\documentclass[aps,prc,twocolumn,nofootinbib,showpacs,floatfix]{revtex4}
\usepackage{epsfig}
\usepackage{bm}
\usepackage{dcolumn}

\def\bea {\begin{eqnarray}}
\def\eea {\end{eqnarray}}
\def\be {\begin{equation}}
\def\ee {\end{equation}}
\def\ben{\begin{enumerate}}
\def\een{\end{enumerate}}
\def\bi{\begin{itemize}}
\def\ei{\end{itemize}}

\def\viz{{\it viz.}\ }

\def\etal{{\it et al.}\ }

 \def\F{{\cal F}}

\def\pr {Phys. Rev.\ }
\def\np {Nucl. Phys.\ }

\def\gAeff{g_{\mbox{\tiny A,eff}}}

\def\gXeff{g_{\mbox{\tiny X,eff}}}

\def\gA{g_{\mbox{\tiny A}}}
\def\gX{g_{\mbox{\tiny X}}}

\def\gL{g_{\mbox{\tiny L}}}
\def\gLeff{g_{\mbox{\tiny L,eff}}}
\def\gS{g_{\mbox{\tiny S}}}
\def\gSeff{g_{\mbox{\tiny S,eff}}}

\def\gP{g_{\mbox{\tiny P}}}
\def\gPeff{g_{\mbox{\tiny P,eff}}}

\def\GF{G_{\mbox{\tiny F}}}

\def\hyphen{{\mbox{-}}}

\def\2p{|2p\rangle }
\def\4p2h{|4p\hyphen 2h\rangle }
\def\6p4h{|6p\hyphen 4h\rangle }
\def\2h{|2h\rangle }
\def\4h2p{|4h\hyphen 2p\rangle }
\def\6h4p{|6h\hyphen 4p\rangle }

\begin{document} 
\preprint{ }
 
\title{A new analysis of $^{14}$O beta decay: branching ratios and CVC consistency}

\author{I.S. Towner}
\altaffiliation{Present address: 
Department of Physics,
Queen's University, Kingston, Ontario K7L 3N6, Canada}
\author{J.C. Hardy}
\affiliation{Cyclotron Institute, Texas A \& M University,                    
College Station, Texas  77843}
\date{\today} 
\begin{abstract} 

The ground-state Gamow-Teller transition in the decay of $^{14}$O is
strongly hindered and the electron spectrum shape deviates markedly 
from the allowed shape.  A reanalysis of the only available data
on this spectrum changes the branching ratio
assigned to this transition by seven standard
deviations: our new result is $(0.54 \pm 0.02)\%$.
The Kurie plot data from two earlier publications are also
examined and a revision to their published branching ratios 
is recommended.
The required nuclear matrix elements are calculated with the shell
model  and, for the first time, consistency is obtained between the M1
matrix element deduced from the analog gamma transition in
$^{14}$N and that deduced from the slope in the shape-correction
function in the beta transition, a requirement of the conserved
vector current hypothesis.  This consistency is only obtained, however,
if renormalized rather than free-nucleon operators are used
in the shell-model calculations.  In the mirror decay of $^{14}$C a 
similar situation occurs.  Consistency between the $^{14}$C lifetime,
the slope of the shape-correction function and the M1 matrix element 
from gamma decay can only be achieved with renormalized operators
in the shell-model calculation.  

\end{abstract} 


\maketitle


\section{Introduction}
\label{s:intro}

The nucleus $^{14}$O decays predominantly by a Fermi transition to the
2.313 MeV $0^+$ state in $^{14}$N.  Weak Gamow-Teller branches are
evident to the $1^+$ state at 3.95 MeV in $^{14}$N (branching ratio, $R_{GT} = 0.055 \%$)
and to the $1^+$ ground state with $R_{GT} \sim 0.6 \%$.  
It can be concluded then that the Fermi
transition has a branching ratio of $\sim 99.3 \%$.
Since systematic studies of Fermi superallowed transitions require
their branching ratios be known to an
accuracy of $\pm 0.1 \%$,  
the branching ratio of the ground-state Gamow-Teller transition
for $^{14}$O must be
determined to within 10\% of its central value.

This Gamow-Teller transition is strongly inhibited.  Its $ft$-value
is roughly $10^4$ times larger than is typical for favoured
$0^+ \rightarrow 1^+$ transitions.  (Even more inhibition
is evident in the analog 
$^{14}$C$~\rightarrow ~ ^{14}$N transition.)  The inhibition is attributed
to accidental cancellation in the allowed Gamow-Teller matrix
element for this transition \cite{RHW68}.  Because the allowed matrix elements
are so small, the induced terms (particularly ``weak magnetism"), 
as well as the
relativistic and the second-forbidden terms are expected to contribute
appreciably to the decay probability.  As a consequence, many of the 
usual assumptions in the allowed approximation may not be valid.
For example, the spectrum shape may deviate markedly from
the allowed (or statistical) spectrum shape.

To date there has only been one measurement, 
by Sidhu and Gerhart \cite{SG66,Si66}, of the detailed shape of the
beta spectrum from $^{14}$O decay, from which they determined the
branching ratio of the
ground-state Gamow-Teller transition with the required
precision.   It was performed with
an iron-free, beta-ray spectrometer, and was published 40 years ago!
More recently, calculations by Garc\'{\i}a and
Brown \cite{GB95} could not satisfactorily fit the observed beta
spectrum,  which led the authors
to suggest there might be some systematic problem
with Sidhu and Gerhart's measurement.
This conclusion would have a serious impact on the branching ratio 
for the Fermi transition.  For this reason, we have         
reanalyzed the data of Sidhu and Gerhart, which we obtained from
a copy of Sidhu's Ph.D. thesis \cite{Si66}.
Our conclusion
is that the $^{14}$O spectrum shape can be understood,
but only if renormalized operators are used in the shell-model
calculations of the nuclear matrix elements.
Our re-analysis yields a ground-state branching ratio of
$(0.54 \pm 0.02) \%$, compared with the originally
published result \cite{SG66} of $(0.61 \pm 0.01) \%$ -- a large
shift in terms of the uncertainties quoted.

\section{Re-analysis of Sidhu-Gerhart experiment}
\label{s:reanal}

The probability per unit time for emission of a positron whose
momentum lies between $p$ and $p + dp$ is
\bea
\frac{dN}{dp} & = & \frac{\GF^2 V_{ud}^2}{2 \pi^3 (\hbar c)^6 \hbar}
p^2 (E_0 - E)^2 , 
\nonumber \\
& &
F(Z,E) S(Z,E) Q(p) R(p) ~ {\rm s}^{-1} ,
\label{dNdp}
\eea
where $\GF$ is the weak interaction coupling constant, $V_{ud}$ the
up-down matrix element of the CKM matrix, $E$ the positron kinetic
energy in MeV units, $E_0$ its maximum value, $p$ the positron
momentum, $F(Z,E)$ the Fermi function, $Z$ the atomic number of
the daughter nucleus, $S(Z,E)$ the shape-correction function, 
$Q(p)$ an atomic
screening correction, and $R(p)$ a radiative correction.
(The average value of $R(p)$, when integrated over an allowed
electron spectrum, is denoted $\delta_R^{\prime}$ in our
publications \cite{TH02} on superallowed Fermi transitions.)
In this work, the shape-correction function includes the
nuclear matrix elements\footnote{Further, these nuclear matrix elements
include the coupling constants in their definition.  For example,
the Gamow-Teller matrix element, $M_{GT}$, includes 
the axial-vector coupling constant $\gA $.}.
We will write
\be
\frac{2 \pi^3 (\hbar c)^6 \hbar \ln 2}{\GF^2 V_{ud}^2
(m_ec^2)^5} = \overline{K} = 6146~{\rm s}
\label{kbar}
\ee
and let all the instrumental factors be encompassed by some
function $K(p)$.  
For an iron-free beta spectrometer, Sidhu assumes the
instrumental function is proportional to the positron momentum,
that is $K(p)=\kappa p$.  
If $dN/dp$ is now interpreted as the
number of counts in the beta-ray spectrometer per unit time that correspond
to positrons whose momentum is between $p$ and $p+dp$,
we can write
\be
\frac{dN}{dp} = A p^3 (E_0 - E)^2 F(Z,E) S(Z,E) Q(p) R(p) ,
\label{countrate}
\ee
where $A = \kappa \ln 2 / \overline{K} (m_e c)^5$ and is a constant
of unknown magnitude.
To calibrate the spectrometer and determine $A$, Sidhu made 
measurements of the lower-energy Fermi transition, for
which all the factors on the right-hand-side of
Eq.~(\ref{countrate}) are calculable, including the shape-correction
function $S(Z,E)$.  In Fig.~22 of his thesis, Sidhu showed the
measured Kurie plot for the Fermi transition defined as
\bea
I(E) \equiv \sqrt{\frac{dN/dp}{p^3 F(Z,E)}} & =  &
A^{1/2} (E_0 - E) 
\nonumber \\
& &
[S(Z,E) 
Q(p) R(p)]^{1/2} .
\label{Kurie}
\eea
In selecting an energy at which $A$ is to be determined,
Sidhu \cite{Si66} writes ``we want to avoid being too close to
the end-point, because of the larger proportion of the high-energy
positrons (from the Gamow-Teller decay) being mixed in this branch
near the end point.  On the other hand, we want to avoid positrons due
to contaminants such as $^{15}$O.  The point at 1.691 MeV should
be a good compromise."  At $E = 1.691$ MeV, 
after reducing his result 
by 1.7\% to eliminate the estimated
contribution of $^{15}$O contaminants,
Sidhu quotes\footnote{
Actually he quotes a value of $1.237 \pm 0.013$, but a trivial
numerical error on page 138 of the thesis indicates that
one entry there is incorrectly recorded.  We have worked
backwards from the published result \cite{SG66} to deduce 
what this value needs to be.  Note the discrepancy is within the
stated error.}
\be
\frac{I^2(E)}{(E_0 - E)^2} = 1.232 \pm 0.013 = A S(Z,E) Q(p) R(p) .
\label{exptIE}
\ee
Without the benefit of today's computing facilties, Sidhu assumed 
the product $S(Z,E) Q(p) R(p)$ to be equal to 2, the square of
Fermi matrix element.  We have now computed these
factors more exactly and obtained:
$S(Z,E) = 1.00226 \times 2.0$,
$Q(p) = 1.00058$ and $R(p) = 0.99523$. This product is $0.2 \%$ less
than Sidhu's choice and yields
\be
A = 0.6173 \pm 0.0064 .
\label{Acalib}
\ee

\begin{table}[ht]
\begin{center}
\caption{Corrected shape-correction functions, $C(Z,W)$, obtained
from Sidhu and Gerhart's data, $J(E)$, via Eq.~(\protect\ref{JEcorr}).   
\label{t:JEcorr}}
\vskip 1mm
\begin{tabular}{cccccc}
\hline
\hline
& & & & & \\
$E$ & $J(E)$ & $F_0 L_0 /F$ & $Q(p)$ & $R(p)$ &
$C(Z,W)$ \\
MeV & Units:$10^{-4}$ & & & & Units:$10^{-4} $ \\[1mm] 
\hline
& & & & & \\
2.235 & 1.738(10) & 1.00296 & 1.00046 & 1.00907 & 2.781(16) \\
2.395 & 1.720(10) & 1.00309 & 1.00043 & 1.00736 & 2.755(16) \\
2.544 & 1.669(10) & 1.00321 & 1.00041 & 1.00571 & 2.679(16) \\
2.702 & 1.652(10) & 1.00334 & 1.00039 & 1.00388 & 2.656(16) \\
2.860 & 1.616(10) & 1.00347 & 1.00037 & 1.00195 & 2.603(16) \\
3.018 & 1.602(10) & 1.00359 & 1.00035 & 0.99987 & 2.585(16) \\
3.172 & 1.553(10) & 1.00372 & 1.00034 & 0.99767 & 2.512(16) \\
3.323 & 1.537(10) & 1.00384 & 1.00033 & 0.99527 & 2.492(16) \\
3.482 & 1.515(10) & 1.00397 & 1.00031 & 0.99239 & 2.463(16) \\
3.640 & 1.500(13) & 1.00410 & 1.00030 & 0.98893 & 2.447(21) \\
3.798 & 1.480(17) & 1.00423 & 1.00029 & 0.98443 & 2.424(28) \\[2mm]
\hline
\hline
\end{tabular}
\end{center}
\end{table}

With $A$ now determined, we consider the higher-energy lower-intensity
Gamow-Teller transition.  For this, we need to perform           
a nuclear-structure calculation of
$S(Z,E)$ and compare with the experimental data.  To this end it
is convenient to introduce the shape-correction function defined by
Behrens and B\"{u}hring \cite{BB82}:
\be
C(Z,W) = \frac{F(Z,E)}{F_0 L_0} S(Z,E) ,
\label{CWdef}
\ee
where $W$ is the total positron energy in electron rest-mass units,
$W = 1 + E/(m_e c^2)$.  The factor $F/F_0 L_0 $ corrects for
the fact that Behrens and B\"{u}hring compute the electron
density at the origin and not at the nuclear radius $R$, as was
historically the case.  In Fig. 20 of his thesis, Sidhu plots
his experimental data for the shape-correction function (corrected for backscattering)
for eleven positron kinetic energies between
2.2 and 3.8 MeV.  Two lower energy points are disregarded 
as being unreliable.  We convert these data 
to the Behrens and B\"uhring-defined 
shape-correction function, and further apply a screening and
radiative correction as follows:
\be
C(Z,W) = \frac{J(E)}{A Q(p) R(p)} 
\frac{F(Z,E)}{F_0 L_0} .
\label{JEcorr}
\ee
Here we have labelled Sidhu's original data as $J(E)$.  
In Table~\ref{t:JEcorr} we give $J(E)$ as well as the 
corrections 
$F/F_0 L_0$, $Q(p)$, $R(p)$ and the determined values of $C(Z,W)$,
which are also plotted in Fig.~\ref{f:CZW}.

\section{Calculation of $C(Z,W)$}
\label{s:CZW}

We will use the formalism of Behrens and B\"{u}hring \cite{BB82} for
the shape-correction function, where it is written in terms
of amplitudes
$M_K(k_e,k_{\nu})$ and
$m_K(k_e,k_{\nu})$:
\bea
C(Z,W) & = &
\sum_{k_e k_{\nu} K} \lambda_{k_e}
\left \{ M_K^2(k_e,k_{\nu}) + m_K^2(k_e,k_{\nu}) \right. 
\nonumber \\
& & \left.
- \frac{2 \mu_{k_e} \gamma_{k_e}}{k_e W}  
M_K(k_e,k_{\nu})  m_K(k_e,k_{\nu}) \right \} . 
\label{CWfinal}
\eea
Here $\lambda_{k_e}$ and $\mu_{k_e}$ are beta-decay Coulomb functions,
which depend on the amplitudes of the electron wave functions at the
origin, and are defined such that their values are of order unity,
with corrections of order $(\alpha Z)^2$.  These quantities have
been tabulated by Behrens and J\"{a}necke \cite{BJ69}.  The
factor $\gamma_{k_e}$ is defined as
$[k_e^2 - (\alpha Z)^2 ]^{1/2}$.  Here $k_e$ and $k_{\nu}$ are
the partial-wave expansion labels for the electron and neutrino
wave functions.  In our evaluations we will keep the lowest
two partial waves in each.  Finally, $K$ is the multipolarity
of the transition operators.  For the ground-state Gamow-Teller
decay, $0^+ \rightarrow 1^+$, the multipolarity is restricted
to $K = 1$ only.

Behrens and B\"{u}hring \cite{BB82} give approximate expressions
for $C(Z,W)$ by expanding the electron wave functions in a
power series in $WR$ and $(\alpha Z)$, and the neutrino wave
function in a power series in $p_{\nu} R$, where $p_{\nu}$
is the neutrino momentum, and $R$ is a typical nuclear size
parameter.  We have not followed this procedure, but rather we have 
computed Eq.~(\ref{CWfinal}) exactly by solving the Dirac
equation for electron wave functions as described in Ref. \cite{HT05}.
Our only simplifying assumption is in the 
evaluation of
$M_K(k_e,k_{\nu})$ and
$m_K(k_e,k_{\nu})$,
where the nuclear form factors of Behrens and B\"{u}hring are 
replaced in the `impulse approximation' by nuclear reduced
matrix elements.  We compute these reduced matrix elements within
the nuclear shell model.

For $A=14$ nuclei, we use the $0p$-shell wave functions of Cohen
and Kurath \cite{CK65} denoted (8-16)POT.  For comparison 
purposes we also consider the more recent $0p$-shell
part of the Warburton-Brown Hamiltonian \cite{WB92} (the
interaction labelled PWBT in Table~X of Ref.~\cite{WB92}).
This interaction was determined from a least squares fit 
to 51 $0p$-shell binding energies for which the rms
deviation of the fit was 378 keV.

These two sets of interactions only incorporate the $0p$-shell.
We wanted also to examine the possible effects of $sd$-shell contributions.
Close to major shell closures the choice of a model space
and effective interaction can be problematic if one
wants to go beyond simple single-major-shell configurations.
For example,
in the $A = 14$ spectrum the lowest-energy states are predominantly
two holes outside a closed $^{16}$O core, $\2h$, but
lying low in the spectrum are `intruder' states with configurations
involving four holes and two particles, $\4h2p$.  Mixing between these
configurations must occur, and to obtain the degree of
mixing with the shell model is difficult.  Shell-model
calculations that attempt to mix $\2h$ and $\4h2p$ configurations
encounter what has been called \cite{WB92} the ``$n \hbar \omega$
catastrophe".  The presence of $\4h2p$ configurations depresses
the $\2h$ states, opening up a large energy gap between the
$\2h$ and $\4h2p$ states.  This would be corrected somewhat if
the model calculation included $\6h4p$ states as well, since the
role of the $\6h4p$ states is to depress the $\4h2p$ states.
Thus when truncating the model space to include only $\2h$ and $\4h2p$
states, the depression driven by the $\6h4p$ states on the $\4h2p$
states is absent.  In an attempt to circumvent this catastrophe
we will use quite weak cross-shell interactions and examine the
sensitivity of our results to the strength of the cross-shell
interaction.

We adopt the following method for incorporating $sd$-shell
effects in the mass-14 system:  We use
the Cohen-Kurath interaction \cite{CK65} for $p$-shell interactions, the
USD \cite{Wi84} for $s,d$-shell interactions and the Millener-Kurath \cite{MK75}
interaction for the cross-shell matrix elements.  The Millener-Kurath
interaction was designed to reproduce the unnatural-parity
states in $p$-shell nuclei, such as the negative-parity states
in $A = 14$ that involve just one particle in the $s,d$-shell.
It wasn't designed to give the mixing between $\2h$ and $\4h2p$
configurations.  Nevertheless we will use the $\langle 2h | V |
4h \hyphen 2p \rangle$ matrix elements given by the 
Millener-Kurath interaction and multiply the matrix elements
by a factor, $f$, that ranges from 0.0 to 0.6.  When
$f = 0.0$, there is no mixing between the $\2h$ and $\4h2p$
configurations, and when $f = 0.6$ the ground-state wave function
is approximately $74 \%$ $\2h$ and $26 \%$ $\4h2p$.  We will
quote results using $f$=0.3, and denote this interaction as MK.  We have
examined the sensitivity of our results to a variation of $f$
and found that the spread of the different results is
within the assigned errors.  

We are interested in further refining the wave function for
the  $1^+~T=0$ state in $A = 14$.  As was
noted by Garc\'{\i}a and Brown \cite{GB95}, who suggested
this procedure, the $GT$ transition strength
to the lowest $1^+$ state is very small compared to the
strength of the transition to the second $1^+$ state
at 3.95 MeV excitation energy.  Thus, any small mixing
between these two $1^+$ states  in the model will have a
large effect on the weak transition rate.  
To make use of this fact, we can write the
wave function
for the lowest $1^+$ state in $^{14}$N as 
\be
| 1^+ {\rm ~low} \rangle = \alpha | 1^+ (1) \rangle
+ \beta | 1^+ (2) \rangle
\label{low}
\ee
with $\alpha^2 + \beta^2 = 1$.  Here $(1)$ and $(2)$ refer
to the first and second model states obtained with
either the CKPOT, PWBT or MK effective Hamiltonians.  In fitting
the beta-decay data, it turns out that  we need a negative sign for $\beta$
with the CKPOT interaction and a positive sign with the
PWBT and MK interactions.

\begin{table*}[ht]
\begin{center}
\caption{The value of the wave 
function amplitude, $\alpha$, Eq.~(\protect\ref{low}) that gave the
best fit to the experimental shape-correction function 
for $^{14}$O, and the best fit to the decay half-life for $^{14}$C.
Also given are the
Gamow-Teller and $M1$ matrix elements for the $0^+ \rightarrow
1^+$ transition, 
the parameters $k$ and $a$ of the shape-correction function, $C(Z,W)$,
and the ground-state branching ratio, $R_{GT}$. 
The $\chi^2$ per degree of freedom, $\chi^2 / \nu$
for the fit to the $^{14}$O shape-correction function is recorded.
Free-nucleon operators were used for the shell-model calculations.
\label{t:alpha}}
\vskip 1mm
\begin{tabular}{rrrrdddr}
\hline
\hline
& & & & & & & \\
 Model~~ & \multicolumn{1}{r}{$\alpha$~~~~} & 
   \multicolumn{1}{r}{$\chi^2 / \nu$} & 
   \multicolumn{1}{r}{$M_{GT}$~~} & 
   \multicolumn{1}{r}{$M_{M1}^{\beta}$~~~~} & 
   \multicolumn{1}{r}{$k$~~~} &
   \multicolumn{1}{c}{$a$} &
   \multicolumn{1}{r}{$R_{GT}(\%)$~} \\
& & & & & & 
    \multicolumn{1}{r}{MeV$^{-1}$} & 
    \\[1mm] 
\hline
& & & & & & & \\
\multicolumn{2}{c}{\bf $^{14}$O decay:} & & & & & & \\[2mm]
CK~~& 0.98447 & 2.0 & 0.01400 & -0.533 & 1.999 &  -0.133 &  0.566 \\
PWBT~~& 0.99805 & 3.5 & 0.01392 & -0.569 & 2.068 & -0.139 & 0.570 \\
MK~~& 0.98482 & 1.5 & 0.01406 & -0.515 & 1.958 & -0.130 & 0.563 \\[5mm] 
\multicolumn{2}{c}{\bf $^{14}$C decay:} & & & & & & \\[2mm]
CK~~& 0.98560 & & $-0.00465$ & 0.503 & 0.345 & -0.647 &  \\
PWBT~~& 0.99760 & & $-0.00469$ & 0.539 & 0.346 & -0.683 &  \\
MK~~& 0.98595 & & $-0.00458$ & 0.484 & 0.354 &  -0.624 &  \\[5mm]
\multicolumn{4}{l}{Expt: $|M_{M1}^{\beta}|$ from $\Gamma_{\gamma}$
in $^{14}$N} & 0.312(7) & & & \\
\multicolumn{3}{l}{Expt:  $^{14}$C slope $a$} & & & & -0.45(4) & \\[2mm]
\hline
\hline
\end{tabular}
\end{center}
\end{table*}

Our strategy is then to adjust $\alpha$ to minimize the $\chi^2$ between the
calculated $C(Z,W)$ and
the corrected experimental shape-correction function given in Table~\ref{t:JEcorr}.
The Gamow-Teller matrix element, $M_{GT}$, is particularly sensitive
to small variations in $\alpha$, and consequently is rather precisely
determined mainly from the fit to the absolute magnitude 
to the shape-correction function.  
We have 
fitted our calculated $C(Z,W)$ to the expression
\be
C(Z,W) = |M_{GT}|^2 k ( 1 + aW + \mu_1 \gamma_1 b/W
+ cW^2 )
\label{CZWexpand}
\ee
as the approximate expressions of Behrens and B\"{u}hring
can be cast in this form.  In Table~\ref{t:alpha} we give the
first two parameters, $k$ and $a$, as determined by least-squares
fitting for each set of shell-model interactions.  The parameter $a$,
which is 
called the slope of the shape-correction function, can be expressed
in approximate form \cite{BB82,GB95} as\footnote{This approximation is 
quite poor.  We only display it to show the important effect of 
the $M1$ matrix element.} 
\be
a_{\rm approx} = \frac{8}{3 M} \frac{M_{M1}^{\beta}}{M_{GT}} ,
\label{aapprox}
\ee
where $M$ is the nucleon mass and
$M_{M1}^{\beta}$ is the $M1$ matrix element.
Clearly this slope is dominated by an interference between
the Gamow-Teller and $M1$ matrix elements.  This is of considerable -- indeed
historical -- importance \cite{GM58}.  A measurement of 
$M_{M1}^{\beta}$ obtained from beta decay 
can be compared with the corresponding matrix element, $M_{M1}^{\gamma}$,
obtained from the electromagnetic 
transition between the isobaric analog state and the ground state of $^{14}$N.
The conserved vector current hypothesis as enunciated by
Gell-Mann \cite{GM58} proposes
that the vector current of the weak interaction 
is just a rotation in isospin of the vector current of
the electromagnetic interaction; in this case we would expect
$M_{M1}^{\beta} = - M_{M1}^{\gamma}/\sqrt{2}$. (The $\sqrt{2}$
factor originates in the isospin Clebsch-Gordan coefficient
for a charge-changing reaction, such as beta-decay, differing
by $\sqrt{2}$ from the isospin Clebsch-Gordan coefficient
for a charge-conserving reaction, such as gamma-decay.)
The analogous
electromagnetic transition in $^{14}$N is the
$0^+~T=1 \rightarrow 1^+~T=0$ decay of the 2.313 MeV state,
which has a measured gamma width \cite{Aj91} of
$\Gamma_{\gamma} = (6.7 \pm 0.3) \times 10^{-9}$ MeV.
The corresponding M1 matrix element is
\be
|M_{M1}^{\gamma}|_{\rm expt} = 
\sqrt{\frac{3 \Gamma_{\gamma}}{4 (E_{\gamma}/\hbar c)^3 \mu_N^2}} = 0.442 \pm 0.010 .
\label{M1expt}
\ee

\begin{table*}[ht]
\begin{center}
\caption{Same as Table~\protect\ref{t:alpha}, except that renormalized
operators are used rather than free-nucleon operators in the
shell-model calculations.
\label{t:renorm}}
\vskip 1mm
\begin{tabular}{rrrrdddr}
\hline
\hline
& & & & & & & \\
Model~~  & \multicolumn{1}{r}{$\alpha$~~~~} & 
   \multicolumn{1}{r}{$\chi^2 / \nu$} & 
   \multicolumn{1}{r}{$M_{GT}$~~} & 
   \multicolumn{1}{r}{$M_{M1}^{\beta}$~~~~} & 
   \multicolumn{1}{r}{$k$~~~} &
   \multicolumn{1}{c}{$a$} &
   \multicolumn{1}{r}{$R_{GT}(\%)$~} \\
& & & & & & 
    \multicolumn{1}{r}{MeV$^{-1}$} & 
    \\[1mm] 
\hline
& & & & & & & \\
\multicolumn{2}{c}{\bf $^{14}$O decay:} & & & & & & \\[2mm]
CK~~& 0.98428 & 5.1 & 0.01490 & -0.313 & 1.552 & -0.099 & 0.540 \\
PWBT~~& 0.99812 & 3.3 & 0.01480 & -0.346 & 1.605 & -0.105 & 0.544 \\
MK~~& 0.98463 & 6.0 & 0.01494 & -0.300 & 1.526 & -0.095 & 0.538 \\[5mm] 
\multicolumn{2}{c}{\bf $^{14}$C decay:} & & & & & & \\[2mm]
CK~~& 0.98556 & & $-0.00338$ & 0.281 & 0.613 & -0.421 &  \\
PWBT~~& 0.99762 & & $-0.00343$ & 0.314 & 0.607 & -0.459 &  \\
MK~~& 0.98590 & & $-0.00333$ & 0.267 & 0.629 & -0.406 &  \\[5mm]
\multicolumn{4}{l}{Expt: $|M_{M1}^{\beta}|$ from $\Gamma_{\gamma}$
in $^{14}$N} & 0.312(7) & & & \\
\multicolumn{3}{l}{Expt:  $^{14}$C slope $a$} & & & & -0.45(4) & \\[2mm]
\hline
\hline
\end{tabular}
\end{center}
\end{table*}

\noindent
Here $E_{\gamma}$ is the photon energy in MeV, $E_{\gamma} = 2.313$ MeV, and
$\mu_N$ the nuclear magneton unit, $\mu_N = 0.1262$ MeV$^{1/2}$ fm$^{3/2}$.
The matrix element is dimensionless.  To conform with CVC,
the beta-decay $M1$ matrix element should be given by
\be
|M_{M1}^{\beta}| = |M_{M1}^{\gamma}|/\sqrt{2} = 0.312 \pm 0.007 ,
\label{M1beta}
\ee
if we neglect
charge-symmetry breaking corrections.
Garc\'{\i}a and Brown \cite{GB95} have studied this issue and
concluded that ``one should not expect very large
charge-symmetry-breaking effects in the $A=14$ system".

It is evident in Table~\ref{t:alpha} that for values of $\alpha$
that give a good fit to the experimental beta-decay spectrum
shape the value of the $M1$ matrix element differs from
Eq.~(\ref{M1beta}) by about a factor of 1.7.  
This disappointing result is also consistent with the conclusions
of Garc\'{\i}a and Brown \cite{GB95}: with $0p$-shell wave
functions it is not possible to fit the $B(M1; 0^+ \rightarrow 1^+)$
radiative decay simultaneously with the beta-decay measurements
of Sidhu and Gerhart \cite{SG66,Si66}.

We have also examined the mirror decay of $^{14}$C, whose lifetime 
is known \cite{NNDC} to be $t_{1/2} = 5700(30)$ yr.  Our strategy 
for this decay is to adjust the wavefunction amplitude $\alpha$ to fit 
this lifetime exactly.  Then the shell-model calculations give a 
prediction for the shape-correction function, $C(Z,W)$, whose slope 
parameter $a$ can be compared with the experimentally determined value of
Wietfeldt \etal \cite{Wi95}.  The results are listed in the bottom 
half of Table~\ref{t:alpha}.  Again we have a disappointing result: 
the calculated slope parameter of $a \simeq -0.65$ MeV$^{-1}$ differs
significantly from the experimental value\footnote{We take the slope 
parameter from Table II of \protect\cite{Wi95} using the fitted value for the 
electron spectrum in the 100 to 160 keV energy range.  In a wider
energy range, $50 - 160$ keV, the authors quote a slope parameter
of $-0.32$ MeV$^{-1}$.} of $-0.45(4)$ MeV$^{-1}$.

\subsection{Renormalized operators}
\label{ss:renorm}

We carried out the
shell-model calculations just discussed using 
operators derived in the impulse approximation with coupling
constants appropriate for free nucleons.  In finite nuclei one expects
corrections to this scheme coming from two sources:  firstly,
the shell-model calculation is carried out in a truncated model space,
which can be corrected in a perturbation expansion,
and secondly the nucleons in the nucleus
are interacting via the exchange of mesons and these mesons can
influence the electromagnetic and weak interactions in nuclei.
The two corrections are called core polarization and
meson-exchange currents respectively.  
These phenomena are responsible for the quenching of the
Gamow-Teller matrix element in finite nuclei.  We define

\be
\gAeff = \gA + \delta \gA ,
\label{gAeff}
\ee
where $\gA$ is the free-nucleon value of the axial-vector
coupling constant, $\gA \simeq 1.27$, and $\delta \gA$ the
correction to it.  We fix the value of $\delta \gA$ by
considering the beta decay of $^{15}$O to its ground-state
mirror in $^{15}$N, whose experimentally determined
Gamow-Teller matrix element \cite{Ra78} shows a reduction
of 13.2(7)\% over that calculated with the free-nucleon
coupling constant for a configuration that consists of a single $0p$-hole
in a closed $^{16}$O core.  To fit the experimental value requires 
\be 
\delta \gA = -0.165 ,
\label{dgA}
\ee
and we will adopt this value.

We also need to understand how the isovector
$M1$ operator is renormalized by core polarization and meson-exchange
currents.  Both these corrections have been evaluated in an 
{\it ab initio} calculation for closed-shell-plus-or-minus-one
nucleon configuration by Towner and Khanna \cite{TK83}.  
Their results for the $A=15$ case of a $0p$ hole in an $^{16}$O core
are expressed in terms of an
effective $M1$ operator defined as
\be
(M1)_{\rm eff} = \gLeff^{(1)} {\bf L} + 
\gSeff^{(1)} {\bf S} + 
\gPeff^{(1)}  [Y_2 , {\bf S} ] ,
\label{M1eff}
\ee
where $\gXeff^{(1)} = \gX^{(1)} + \delta \gX^{(1)}$, with $\gX$ being the free-nucleon
coupling constant, $\delta \gX$ the calculated correction
to it and $X = L$, $S$ or $P$.  The free-nucleon values are
$\gL^{(1)} = 0.5$, $\gS^{(1)} = 4.706$ and $\gP^{(1)} = 0$.  
Towner and Khanna calculated
the isovector
combinations of the magnetic moments of the $^{15}$O and
$^{15}$N ground states and obtained an $8.9 \%$
enhancement in the free-nucleon isovector magnetic moment compared
to an experimental enhancement of $11.1 \%$ -- a clear success for an
{\it ab initio} calculation.  
We, therefore, adopt their calculated values of
\be
\delta \gL^{(1)} = 0.076 ~~~~~ \delta \gS^{(1)} = -0.22 
 ~~~~~ \delta \gP^{(1)} = 0.96 .
\label{dgM1}
\ee

\begin{figure}[t]
\vspace{-2.0cm}
\epsfig{file=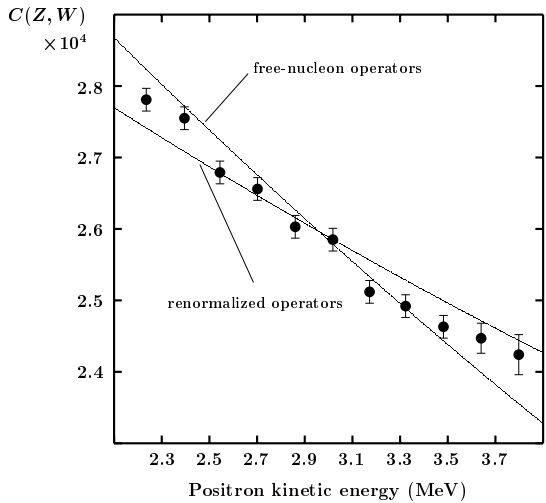,width=8.5cm}
\vspace{-3.0cm}
\caption{The shape-correction function calculated with PWBT wave functions  
compared with the experimental data of Sidhu and Gerhart \protect\cite{Si66}.
\label{f:CZW}}
\end{figure}

With the effective operators thus selected, we repeated the 
strategy of adjusting 
the wave function amplitude $\alpha$ to minimize the $\chi^2$ 
between the calculated $C(Z,W)$ and the corrected experimental 
shape-correction function given in Table~\ref{t:JEcorr}.
The results of the fit are given in 
Table~\ref{t:renorm}, where it is observed that the $M_{M1}^{\beta}$ 
matrix element
is considerably reduced from that in Table~\ref{t:alpha} and is now
comparable to that deduced from gamma decay.  This is a major
success and a strong endorsement for the use of renormalized 
operators.  However, it is not all good news.  For two of the
wave function choices, CK and MK, the quality of the fit to
the experimental shape-correction function is inferior, the
$\chi^2$ per degree of freedom being two to four times larger.
Only for the PWBT interaction is the quality of the fit comparable.
We show the PWBT fit in Fig.~\ref{f:CZW}.  It is clear that the
renormalized operators lead to a smaller slope in the
shape-correction function.  This is the expected result since
the slope is governed by the matrix-element ratio
$M_{M1}^{\beta} / M_{GT}$, as given in Eq.~(\ref{aapprox}). 

For the mirror decay in $^{14}$C, our strategy was to fix
the wavefunction amplitude $\alpha$ to reproduce the known
lifetime, and use these wave functions to compute
the shape-correction function $C(Z,W)$, giving a prediction
for the slope parameter, $ a$.  From Table~\ref{t:renorm} 
it is clear that the use of renormalized operators is very
successful:  the calculated slope parameter of
$a \simeq -0.43$ MeV$^{-1}$ agrees perfectly with the
experimental result of Ref.~\cite{Wi95}.

\section{The $^{14}$O Gamow-Teller branching ratio}
\label{s:RGT}

\begin{table}[ht]
\begin{center}
\caption{Statistical rate function $f$, Eq.~(\protect\ref{fexact}),
for the Gamow-Teller transitions in $^{14}$O and $^{14}$C decays.
\label{t:fvalu}}
\vskip 1mm
\begin{tabular}{rrr}
\hline
\hline
& & \\[-2mm]
& \multicolumn{2}{c}{$f$} \\[2mm]
\cline{2-3}
& & \\[-2mm]
 Model~~ & \multicolumn{1}{c}{$^{14}$O} & 
   \multicolumn{1}{c}{$^{14}$C} \\[1mm]
\hline
& & \\
\multicolumn{3}{l}{\bf free-nucleon operators}  \\[2mm]
CK~~ & 2473.1 & $1.5552 \times 10^{-3}$ \\
PWBT~~ & 2522.9 & $1.5284 \times 10^{-3}$ \\
MK~~ & 2441.8 & $1.6040 \times 10^{-3}$ \\[5mm]
\multicolumn{3}{l}{\bf renormalized operators}  \\[2mm]
CK~~ & 2086.4 & $2.9314 \times 10^{-3}$ \\
PWBT~~ & 2128.1 & $2.8437 \times 10^{-3}$ \\
MK~~ & 2067.5 & $3.0222 \times 10^{-3}$ \\[5mm]
$f_{\rm stat}$~~ & 1633.6 & $6.0766 \times 10^{-3}$ \\[2mm]
\hline
\hline
\end{tabular}
\end{center}
\end{table}

Now that a shape-correction function $C(Z,W)$ has been obtained
for $^{14}$O decay that agrees reasonably well with the data of 
Sidhu and Gerhart \cite{SG66,Si66} (see Fig.~\ref{f:CZW}) 
we integrate this function
over the entire beta spectrum
and compare with the analogous Fermi transition
to obtain the Gamow-Teller branching ratio, $R_{GT}$:
\bea
R & = & \frac{t_F}{t_{GT}} = \frac{N_{GT}}{N_F} =
\frac{f_{GT} (1 + \delta_R^{\prime})_{GT} |M_{GT}|^2}  
{f_{F} (1 + \delta_R^{\prime})_{F} |M_{F}|^2 (1 - \delta_C )} , 
\nonumber \\
R_{GT} & = & \frac{R}{1 + R} ,
\label{Ratio}
\eea
where $t_F$ and $t_{GT}$ are the partial half-lives of the
Fermi and Gamow-Teller transitions respectively, 
and $N_F$ and $N_{GT}$ are their integrated count rates,
$N = \int ( dN/dp)dp$.  For the Gamow-Teller transition, the
statistical rate function 
is defined as
\be
f = \frac{1}{|M_{GT}|^2} \int_1^{W_0} p W
(W_0 - W)^2 F_0 L_0 C(Z,W) Q(p) dW .
\label{fexact}
\ee
Note that, since $C(Z,W)$ includes the nuclear matrix elements,
we have divided by $|M_{GT}|^2$ to
conform with the normal definition of $f$. The calculated
values for the Gamow-Teller transition are listed in
Table~\ref{t:fvalu} for the various wave function selections. 

The other factors in Eq.~(\ref{Ratio}) were evaluated as follows:
For the Fermi
transition we calculate $f_F = 42.772$ (see Ref.~\cite{HT05}) and $|M_F|^2 = 2.0$.
Next, using the methods described in Ref.~\cite{TH02} we obtain
$(\delta_R^{\prime})_{GT} = 1.294 \%$,   
$(\delta_R^{\prime})_{F} = 1.520 \%$ and   
$\delta_C = 0.57 \%$.  With these values we obtain the
branching ratios 
listed in the last column of Tables~\ref{t:alpha}
and \ref{t:renorm}.
We adopt the result with renormalized operators as our
central value and assign an error that is half the spread between
the renormalized and free-nucleon operators, obtaining
$R_{GT} = ( 0.540 \pm 0.015 ) \%$ .
However, there still remains a $1 \%$ normalization uncertainty
in the value of the calibration
constant $A$, Eq.~(\ref{Acalib}).  We add this uncertainty linearly to
obtain as our final branching ratio a value of
\be
R_{GT} = ( 0.54 \pm 0.02 ) \% .
\label{Rfinal}
\ee
This result differs from Sidhu and Gerhart's \cite{SG66} published result
of $R_{GT} = (0.61 \pm 0.01 ) \%$ and has a larger uncertainty.
The Gamow-Teller matrix element from
the fit is $M_{GT} = 0.0149 \pm 0.0005$ compared to Sidhu and
Gerhart's published result of 
$M_{GT} = 0.0164 \pm 0.0004$.

\section{Impact on Superallowed branch}
\label{s:impact}

\subsection{Two earlier measurements}
\label{ss:early}

There are two earlier measurements of the $^{14}$O branching ratio:  
Sherr \etal \cite{Sh55} obtained $R_{GT} = (0.6 \pm 0.1 ) \%$,
while Frick \etal \cite{Fr63} obtained $(0.65 \pm 0.05) \%$.  In
both cases Kurie plots were constructed from raw data for both the
Fermi and Gamow-Teller transitions and the required
branching ratio was obtained from their ratio:
\be
R = \frac{f_{GT}}{f_{F}} \frac{X_{GT}^2}{X_{F}^2} ~~~~~~ R_{GT} = \frac{R}{1+R},
\label{RKurie}
\ee
where $X_F$, $X_{GT}$ are the ratio of the Kurie plot data to the
allowed approximation $(W_0 - W)$, \viz
\be
X = \frac{1}{n_i} \sum_{i = 1}^{n_i} \frac{K(W_i)}{W_0 - W_i} ~~~~~~
K(W) = \sqrt{\frac{dN/dp}{p^2 F(Z,W)}} .
\label{Xallow}
\ee
Here $W_i$ are the values of $W$ for which experimental data have
been obtained, and $n_i$ is the number of such data.
The unknown normalization of the Kurie plots cancels in the ratio.
Further, $f$ is the integrated electron spectrum, which in the
allowed approximation is 
\be
f_{\rm stat} = \int_1^{W_0} p W (W_0 - W)^2 F(Z,W) Q(p) dW .
\label{fstat}
\ee
Sherr \etal \cite{Sh55} only published their Gamow-Teller Kurie plot,
so it is not possible to reanalyze their result.  Frick \etal \cite{Fr63},
on the other hand, published both their Fermi and Gamow-Teller Kurie plots, 
so we have reanalyzed them according to Eq.~(\ref{RKurie}) obtaining
$R_{GT} = (0.64 \pm 0.03) \%$ agreeing satisfactorily with the
published result.  (Our uncertainty only includes the statistical uncertainty
in the fit of the allowed shape to the Kurie data, and the
uncertainty in the end point energy, $W_0$, as known in 1963).

The original analyses of both experiments were
based on the allowed approximation.  However, as we have
discussed at length, the $^{14}$O Gamow-Teller transition is strongly
hindered and the allowed approximation not sufficient.  Thus, we have reanalyzed 
the Frick \etal \cite{Fr63} Kurie plot data using for $X$
\be
X = \frac{1}{n_i} \sum_{i = 1}^{n_i} \frac{K(W_i) M_{GT}}{(W_0 - W_i) \sqrt{C(Z,W_i)}},
\label{Xexact}
\ee
and using for the integrated spectrum, $f$, the exact expression given
in Eq.~(\ref{fexact}).  The shape-correction function, $C(Z,W)$, is
evaluated with $p$-shell wave functions, PWBT, and renormalized
operators.  The result is an increase in the branching ratio
to $R_{GT} = (0.73 \pm 0.03) \%$.  We checked our procedure by
performing the same analysis on Sidhu's Kurie plots, as published
in his thesis \cite{Si66}.  In the allowed approximation we
obtained $R_{GT} = ( 0.45 \pm 0.01) \%$, while with the
exact expressions we obtained $R_{GT} = (0.53 \pm 0.01) \%$ in
agreement with our more accurate analysis of the shape-correction
functions.  The conclusion is clear.  For the Gamow-Teller 
branching ratio in $^{14}$O, determinations based on an
allowed approximation analysis of Kurie plots  have to be increased 
by about 14\%.  Unfortunately this places the earlier data in conflict 
with the more accurate Sidhu-Gerhart result \cite{SG66,Si66}.  This
is not a question of the method of analysis: the raw data are in
conflict.  If one considers a ratio of ratios, comparing the
ratio of Gamow-Teller to Fermi Kurie plots for Sidhu \cite{Si66}
and Frick \etal \cite{Fr63}, discrepacies of order 20\% are
evident.  This, alone, leads to a 40\% difference in the
deduced branching ratios.  If a modern day experiment were
to be mounted, this would be one discrepancy that could quickly
be resolved.

\subsection{The Fermi branch and corrected $\F t$ value}
\label{ss:fermi}

A survey of all the data on superallowed $0^+ \rightarrow 0^+$ Fermi
decay has recently been published by Hardy and Towner \cite{HT05}. 
For the $^{14}$O Fermi branching ratio, the value of 99.334(10)\% is given there
based on the ground-state Gamow-Teller 
branching ratio obtained from Sidhu and Gerhart \cite{SG66}
averaged with the two older and less precise results from \cite{Sh55}
 and \cite{Fr63}, and a second Gamow-Teller branching ratio to the 3.95 MeV 
state of 0.0545(19)\%.  If we now replace the Sidhu-Gerhart value with
the result from Eq.~(\ref{Rfinal}),  and increase the branching
for the two older data by 14\% as discussed in
Sect.~\ref{ss:early} but leave their error assignments
at their published values, then we obtain a
very conservative estimate of the ground-state
Gamow-Teller branching ratio of $(0.57 \pm 0.06)\%$.
The uncertainty here has been scaled by 2.6 
acoording to our usual prescription \cite{HT05} because of the incompatility 
of the Sidhu-Gerhart result with the two 
older measurements.  The Fermi branching ratio 
is now increased to 99.376(65)\%.  The impact of this is to lower 
the corrected $\F t$ value from 3071.9(26)s to 3070.7(32)s.  This 
value still leaves the $^{14}$O datum consistent with the other eleven
precision measured $\F t$ values, 
although it is now on the low side of the average.  The slight shift in 
$^{14}$O is well within the stated errors in the survey \cite{HT05} 
and has a negligible impact on the physics conclusions obtained there.

\section{A note on $ft$ values for Gamow-Teller transitions}
\label{s:ft}

It is traditional to characterize an allowed Gamow-Teller transition
by its $\log ft$ value, where the statistical rate function $f$
used in this application is devoid of any nuclear-structure factors and
is defined in Eq.~(\ref{fstat}).
When the Gamow-Teller matrix element is large, of order unity,
then the shape-correction function $C(Z,W)$ is nearly
independent of energy and $C(Z,W)/|M_{GT}|^2$ is
close to unity. Under these conditions there is little difference between the
exactly defined $f$ of Eq.~(\ref{fexact}) and the
traditional expression in Eq.~(\ref{fstat}).  But for the
Gamow-Teller transitions in $^{14}$O and $^{14}$C decays,
$|M_{GT}|$ is very small and the shape-correction function has
a significant effect.  For these transitions
there is a large difference between the
exact $f$ and $f_{\rm stat}$ as shown in Table~\ref{t:fvalu}.

For the exactly-defined $f$, the transition $ft$-value equals
a constant divided by the square of the Gamow-Teller matrix element
\be
ft = \frac{2 \pi^3 \ln 2 (\hbar c)^6 \hbar }{\GF^2 V_{ud}^2
(m_e c^2)^5} \frac{1}{|M_{GT}|^2} = \frac{6146}{|M_{GT}|^2}~{\rm s} .
\label{ft}
\ee
Note that in our notation $|M_{GT}|$ includes the axial-vector
coupling constant, $\gA$. 
For precision work, the lifetime $t$ should 
be adjusted for radiative corrections.  Eq.~(\ref{ft}) is
frequently used to deduce the Gamow-Teller matrix element from
a published $ft$ value for which $f_{\rm stat}$ has been used
for $f$.  For example, the National Nuclear Data Center \cite{NNDC}
gives the $\log ft$ value for $^{14}$C decay as 9.040(3) and
for $^{14}$O decay as 7.279(8).  To deduce $|M_{GT}|$ from Eq.~(\ref{ft}) with 
these values would be incorrect.  The $\log ft$ values
for the exactly defined $f$ are 8.72(2) and 7.44(1) respectively,
with larger error bars because of the uncertainty from nuclear
structure.

\section{Conclusions}
\label{s:concl}

We began this work disturbed by the statement from Garc\'{\i}a and
Brown \cite{GB95} that it was not possible to fit the
$B(M1;0^+ \rightarrow 1^+)$ radiative decay in $^{14}$N
simultaneously with the $^{14}$O beta decay measurements of
Sidhu and Gerhart \cite{SG66}, an apparent violation of the conserved
vector current hypothesis.  A problem with the Sidhu-Gerhart work
``could significantly change the conclusions extracted from
$0^+ \rightarrow 0^+$ transitions regarding universality and
unitarity", they wrote.  We initially reanalyzed the
Sidhu-Gerhart experiment and came to the same conclusion as
Garc\'{\i}a and Brown, but we then discovered that by using 
renormalized operators in the
shell-model calculation we could achieve much greater consistency 
between the requirements   of CVC and the measurements of
Sidhu and Gerhart.  This observation was reinforced when we examined
the mirror $^{14}$C decay.  There it was only possible to fit the known
lifetime and the slope parameter in the shape-correction function \cite{Wi95}
while remaining consistent
with the requirements of CVC, if renormalized operators
were used.  This is our principal physics conclusion:  the use of
renormalized operators is mandatory.

A second important outcome relates to the superallowed beta decay of $^{14}$O.
With the reanalysis of the Sidhu-Gerhart experiment the recommended 
value from this experiment for the Gamow-Teller branching ratio 
from $^{14}$O to the ground state
of $^{14}$N  
is $R_{GT} = (0.54 \pm 0.02)\%$
compared to the published value \cite{SG66} of $(0.61 \pm 0.01)\%$.
This result,
when combined with updated older measurements,
 revises the recommended 
branching ratio for the Fermi transition from 99.334(10)\% 
to 99.376(65)\% and shifts the corrected $\F t$ value for the Fermi
branch downwards by 1.2s.  This shift is within the stated 
uncertainty of the $\F t$ value given in the survey of Hardy
and Towner \cite{HT05} and does not alter any of the conclusions reached
there.



\end{document}